\documentclass[twocolumn,showpacs,preprintnumbers,amsmath,amssymb]{revtex4}
\usepackage{graphics}% Include figure files
\usepackage{epsfig}
\usepackage{bm}% bold math
\begin{document}
%\preprint{Krenke01}
\title{Magnetization easy-axis in martensitic Heusler alloys estimated by strain measurements under magnetic-field}
\author{Seda Aksoy, Thorsten Krenke\cite{AdressKrenke}, Mehmet Acet, Eberhard F. Wassermann}

\affiliation{Experimentalphysik, Universit\"{a}t Duisburg-Essen,
D-47048 Duisburg, Germany}

\author{Xavier Moya, Llu\'{i}s Ma\~nosa, Antoni Planes}
\affiliation{Facultat de F\'isica, Departament d'Estructura i
Constituents de la Mat\`eria, Universitat de Barcelona, Diagonal
647, E-08028 Barcelona, Catalonia, Spain}
\date{\today}

\begin{abstract}
We study the temperature dependence of strain under constant
magnetic-fields in Ni-Mn based ferromagnetic Heusler alloys in the
form Ni-Mn-$X$ ($X$: Ga, In, Sn, Sb) which undergo a martensitic
transformation. We discuss the influence of the applied
magnetic-field on the nucleation of ferromagnetic martensite and
extract information on the easy-axis of magnetization in the
martensitic state.
\end{abstract}


\maketitle

At off-stoichiometric compositions with respect to the 2-1-1
stoichiometry, almost any Ni-Mn based ferromagnetic Heusler alloy
in the form Ni-Mn-$X$($X$: group IIIA-VA elements) undergoes a
martensitic transformation
\cite{Soderberg06,Sutou04,Thorsten06,Thorsten05}. Ni$_2$MnGa is
the only Ni-Mn-based Heusler that undergoes such a transition in
the 2-1-1 stoichiometry \cite{Ullakko96}. This occurs at roughly
240 K, and it exhibits large magnetic-field-induced strains in the
martensitic state. Many of the off-stoichiometric alloys with $X$
as In, Sn, Sb, etc., exhibit a magnetic-field-induced reverse
martensitic transformation from the martensitic to the austenitic
state and is usually accompanied by large strains
\cite{Sozinov02,Oikawa06,Thorsten07b}and magnetocaloric effects
\cite{Krenke05,Moya07,Khan07}. These effects are closely related
to the strength of the magnetoelastic coupling in the martensitic
state.

A knowledge of the magneto-crystalline anisotropy is particularly
important to be able to understand the subtleties associated with
the coupling between the magnetic and structural degrees of
freedom, this being responsible for large strains and entropy
changes around the transition temperature
\cite{OHandley98,Albertini01}. Anisotropy studies usually require
single crystals of bulk or thin film specimens on which various
experimental techniques such as neutron diffraction, ferromagnetic
resonance, magnetic circular dichroism, torque studies, etc. can
be performed. These can provide information on the easy-axis of
magnetization and on the orbital and spin moments, which are prime
parameters for understanding the magnetic anisotropy, and,
therefore, the magneto-crystalline anisotropy.

In the L1$_0$ phase, having two $a$-axes and a $c$-axis, or in any
modulated martensitic structure, the magnetization will tend to
lie either in a plane bounded by the $a$-axes or along the
$c$-axis of the unit-cell. We demonstrate in this study that by
using polycrystalline specimens, it is possible to provide
information on the easy-axis of magnetization in the martensitic
structure with temperature-dependent strain measurements under
constant magnetic-field. We present results on such measurements
for the martensitic Heusler alloys Ni-Mn-Sn, Ni-Mn-In, and
Ni-Mn-Sb; as well as Ni-Mn-Ga, which serves as a reference system
for which the easy-axis of magnetization in the martensitic state
is known to be along the short-axis \cite{Ullakko96}.

\begin{figure}
\includegraphics[width=8cm]{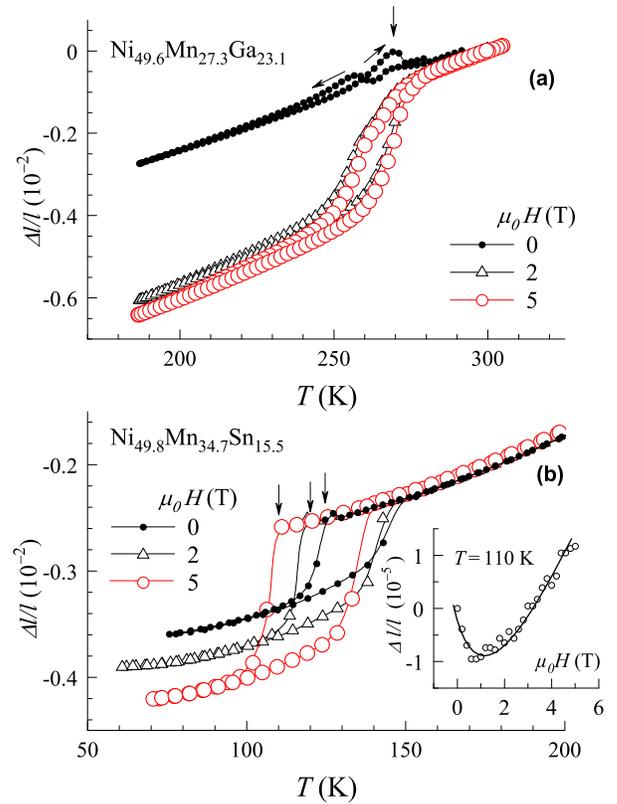}
\caption{\label{strain1} (color online) color online) $\Delta l/l$
versus temperature under 0, 2, and 5 T for (a)
Ni$_{49.6}$Mn$_{27.3}$Ga$_{23.1}$ and (b)
Ni$_{49.8}$Mn$_{34.7}$Sn$_{15.5}$.  Vertical arrows indicate to
$M_s$ Inset in part (b) shows the field-induced strain after
cooling the sample through $M_s$ to 110 K in zero-field.}
\end{figure}

Ingots were prepared by arc melting pure metals under argon
atmosphere in a water cooled Cu crucible. They were then
encapsulated under argon atmosphere in quartz glass, annealed at
1073 K for 2 hours and, afterwards, quenched in ice-water. The
compositions of the alloys were determined by energy dispersive
x-ray analysis. Temperature-dependent strain measurements in
magnetic-fields up to 5 T were carried out using strain gauges.
The field was first applied in the austenitic state, and the data
were subsequently taken on decreasing and then on increasing
temperature. The strain was measured parallel to the applied
field.

The relative length-change $\Delta l/l$ referenced to 300 K and in
magnetic-fields up to 5 T for Ni$_{49.6}$Mn$_{27.3}$Ga$_{23.1}$ is
shown in Fig. \ref{strain1}(a). In the absence of a
magnetic-field, a weak hysteretic feature is found in the
temperature range corresponding to that of the martensitic
transition, and no substantial difference in the macroscopic
dimensions between the austenitic and martensitic states is found.
Applying a magnetic-field in the austenitic state has very little
influence on the strain in this state. However, when the sample is
subsequently cooled through the martensitic transition temperature
$M_s$, a large difference in the strain between the austenite and
martensite states occurs. This effect has previously been observed
in a single-crystal specimen and is due to the alignment of the
short easy-axis of magnetization ($c-$axis) with the external
field, causing the sample to shrink along the applied field
direction \cite{Ullakko96}. With increasing measuring field, the
difference in strain increases due to increased preferred
alignment of the short $c-$axis along the applied field. High
twin-boundary mobility in Ni-Mn-Ga is expected to be the main
cause of the alignment, although martensite variant nucleation
with preferred $c-$axis orientation in the external field already
just at $M_s$ is also an influence on the shrinkage.

Ni$_{49.8}$Mn$_{34.7}$Sn$_{15.5}$ undergoes a martensitic
transformation at about 120 K. As seen in Fig. \ref{strain1}(b),
cooling in the absence of a magnetic-field leads to a raid drop in
$\Delta l/l$ at $M_s$ indicating a volume decrease. The presence
of this volume decrease is sustained by temperature-dependent
neutron diffraction experiments \cite{Brown06}. Cooling in the
presence of a magnetic-field causes $M_s$ (indicated by arrows) to
drop at a rate of about $-3$ KT$^{-1}$. At the same time, the
difference in strain between the austenite and martensite states
increases with increasing magnetic-field as in the case of the
data of the Ni-Mn-Ga sample in Fig. \ref{strain1}(a). However,
twin boundary mobility in Ni-Mn-Sn sample is weak, so that only
little magnetic-field-induced strain ($\sim 10^{-5})$ is observed
in fields up to 5 T after cooling the sample through $M_s$ in
zero-field to 110 K (inset Fig. \ref{strain1}(b)). Therefore, the
large change in strain ($\sim 10^{-3})$ between the austenite and
martensite states in this material should be related to the effect
of the magnetic-field in providing a preferred orientation for the
martensite  variants during their nucleation. Since the strain
difference increases with increasing magnetic-field (meaning that
the sample-length decreases with increasing cooling-field), the
easy-axis of magnetization in the martensitic state is expected to
be along a short-axis or at least in a plane bounded by the
short-axes.

\begin{figure}
\includegraphics[width=8cm]{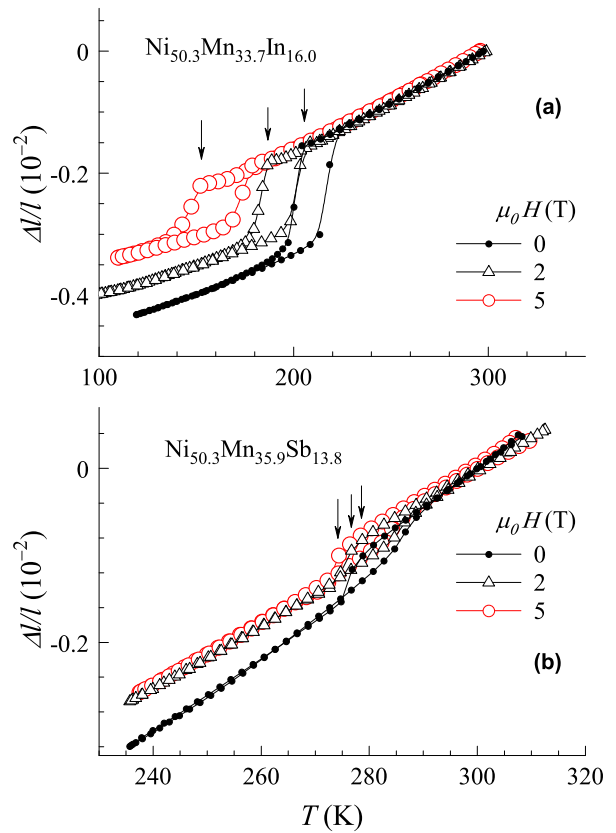}
\caption{\label{strain2} (color online) $\Delta l/l$ versus
temperature under 0, 2, and 5 T for (a)
Ni$_{50.3}$Mn$_{33.7}$In$_{16.0}$ and (b)
Ni$_{50.3}$Mn$_{35.9}$Sb$_{13.8}$. Vertical arrows indicate to
$M_s$.}
\end{figure}

\begin{figure}
\includegraphics[width=8cm]{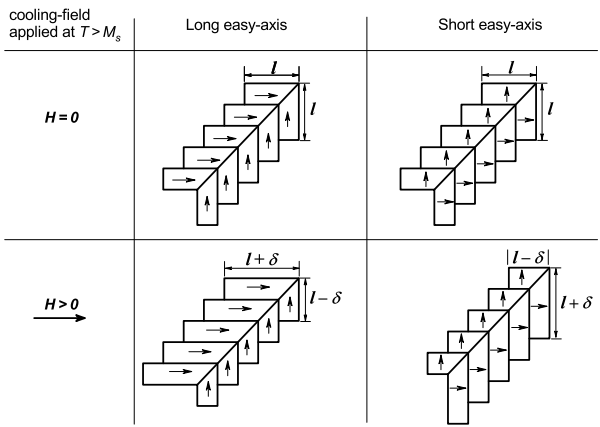}
\caption{\label{strain3} Schematic representation of ferromagnetic
martensite nucleation with and without a cooling magnetic-field
applied at $T>M_s$. Twins are represented with tetragonal units of
length $l$ built up of self-similar tetragonal unit-cells. There
is no preferred variant growth during martensite nucleation when
cooled in $H=0$. Preferred variant growth during martensite
nucleation occurs when the sample is cooled through $M_s$ in
$H>0$, such that when the long-axis is the easy-axis, the length
increases in field direction by $\delta$. When the short-axis is
the easy-axis the length decreases by $\delta$ in the field
direction.}
\end{figure}

The situation for Ni$_{50.3}$Mn$_{33.7}$In$_{16.0}$ is the
opposite: As seen in Fig. \ref{strain2}(a), the cooling-field acts
to decrease the relative length-change between the austenite and
the martensite states. This can happen if the long-axis of a
martensite variant is preferred along the field direction, which
would indicate that a long-axis is the easy-axis of magnetization.
The rate of change of $M_s$ with applied field for this sample is
about $-10$ KT$^{-1}$, being about 3 times larger than that for
the Ni-Mn-Sn sample.

The data for Ni$_{50.3}$Mn$_{35.9}$Sb$_{13.8}$ is similar to that
of Ni$_{50.3}$Mn$_{33.7}$In$_{16.0}$ Fig. \ref{strain2}(b). Here,
the rate of decrease of $M_s$ with applied field is about $-1$
KT$^{-1}$. The relative length-change between the austenite and
martensite states in applied field also decreases with increasing
applied field but, above 2 T further decrease becomes
insignificant. In this case also, the easy-axis is expected to be
a long-axis.

For the measurements of all samples presented above, the data are
reproducible on cycling through the transition temperatures under
zero-field or under the presence of a field. Microscopic morphological
modifications such as cracks or other deformations, which could occur during
cycling and lead to irreproducibilities in the strain-data, are not found.

It is known that when a martensitic material is cooled through
$M_s$ under zero applied field, the martensitic variants
self-organize forming twin-related structures in order to minimize
the elastic energy associated with the change of shape of the
unit-cell. This minimizes the macroscopic deformation due to the
structural transformation. The data presented here suggests that
when cooling the system through $M_s$ under an applied
magnetic-field, martensite variants can nucleate and grow with a
preferred orientation within the austenitic matrix. The particular
orientation would depend on the direction of the applied field and
the direction of the easy-axis of magnetization in the martensitic
phase. In this case, considerable modifications in the macroscopic
strain can be obtained with respect to the case where the sample
is cooled in the absence of a field. This nucleation mechanism
does not depend on the mobility of the martensitic variants.

As a summary, we show schematically in Fig. \ref{strain3} the
effect of a cooling magnetic-field on martensite nucleation probed
through strain. In this figure, 90$^\text {o}$ twins are
represented with tetragonal units of length $l$ built up of
self-similar tetragonal unit-cells. When a sample is cooled
through $M_s$ in the absence of a field, no preferred direction is
given to variant-growth during nucleation, whether the easy-axis
of magnetization is a long-axis or a short-axis. When a field is
applied in the austenitic state and the sample is cooled through
$M_s$ in this field, a preferred growth-direction is provided to
the variants. Variants with easy-axis along the applied field
direction nucleate more. If the easy-axis is the long axis, the
sample-length measured along the field direction increases by an
amount to $l+\delta$. If the easy-axis is the short-axis, the
sample-length decreases to $l-\delta$. In this manner, it is
possible to gain an idea on the easy-axis of magnetization in the
martensitic state using polycrystalline specimens, and further
studies on anisotropy properties can be carried out using such
results.

This work was supported by Deutsche Forschungsgemeinschaft
(SPP1239) and CICyT (Spain), project MAT2007-61200. XM
acknowledges support from DGICyT (Spain).

\end{document}